# Angular momentum in rotating superfluid droplets

Sean M. O'Connell[1], Rico Mayro P. Tanyag[1,2], Deepak Verma[1], Charles Bernando[3,4], Weiwu Pang[5], Camila Bacellar[6,7], Catherine A. Saladrigas[6,7], Johannes Mahl[6,8], Benjamin W. Toulson[6], Yoshiaki Kumagai[9], Peter Walter[10], Francesco Ancilotto[11,12,a)], Manuel Barranco[13,14,15], Marti Pi[13,14], Christoph Bostedt[9,16,17,18,19], Oliver Gessner[6,a)] and Andrey F. Vilesov[1,3,a)]

[1] Department of Chemistry, University of Southern California, Los Angeles, California 90089, USA
[2] Technische Universät Berlin, Institut für Optik und Atomare Physik, 10623 Berlin, Germany
[3] Department of Physics and Astronomy, University of Southern California, Los Angeles, California 90089, USA
[4] OVO (PT. Visionet Internasional), Lippo Kuningan 20th floor, Jl. HR Rasuna Said Kav. B-12, Setiabudi, Jakarta, 12940, Indonesia
[5] Department of Computer Science, University of Southern California, Los Angeles, California 90089, USA
[6] Chemical Sciences Division, Lawrence Berkeley National Laboratory, Berkeley, California 94720, USA
[7] Department of Chemistry, University of California Berkeley, Berkeley, California 94720, USA
[8] Department of Physics, University of Hamburg, 22761 Hamburg, Germany
[9] Argonne National Laboratory, 9700 South Cass Avenue B109, Lemont, Illinois 60439, USA
[10] Linac Coherent Light Source, SLAC National Accelerator Laboratory, Menlo Park, California, 94025, USA
[11] Dipartimento di Fisica e Astronomia and CNISM, Università di Padova, 35122 Padova, Italy
[12] CNR-IOM Democritos, 34136 Trieste, Italy
[13] Departament FQA, Universitat de Barcelona, Facultat de Física, 08028 Barcelona, Spain
[14] Institute of Nanoscience and Nanotechnology (IN2UB), Universitat de Barcelona, Barcelona, Spain
[15] Laboratoire des Collisions, Agrégats et Réactivité, IRSAMC, Université Toulouse 3, F-31062 Toulouse, France
[16] Department of Physics and Astronomy, Northwestern University, 2145 Sheridan Road, Evanston, Illinois 60208, USA
[17] Paul Scherrer Institut, 5232 Villigen – PSI, Switzerland
[18] LUXS Laboratory for Ultrafast X-ray Sciences, Institute of Chemical Sciences and Engineering, Lausanne, Switzerland
[19] École Polytechnique Fédérale de Lausanne (EPFL), CH-1015, Lausanne, Switzerland

a) Authors to whom correspondence should be addressed. Electronic addresses: francesco.ancilotto@pd.infn.it; ogessner@lbl.gov; and vilesov@usc.edu.

**Abstract**

The angular momentum of rotating superfluid droplets originates from quantized vortices and capillary waves, the interplay between which remains to be uncovered. Here, the rotation of isolated sub-micrometer superfluid $^4$He droplets is studied by ultrafast x-ray diffraction using a free electron laser. The diffraction patterns provide simultaneous access to the morphology of the droplets and the vortex arrays they host. In capsule-shaped droplets, vortices form a distorted triangular lattice, whereas they arrange along elliptical contours in ellipsoidal droplets. The combined action of vortices and capillary waves results in droplet shapes close to those of classical droplets rotating with the same angular velocity. The findings are corroborated by density functional theory calculations describing the velocity fields and shape deformations of a rotating superfluid cylinder.

A quiescent droplet is spherical, whereas it experiences pronounced shape deformations when set into rotation. Classical viscous droplets execute rigid body rotation (RBR), the kinematics of which are well understood [1-4]. In contrast, a superfluid droplet is described by a quantum mechanical wave function [5,6] and, at low temperatures, its angular momentum may be stored in quantized vortices as well as capillary waves. The rotation of superfluid droplets poses intriguing questions regarding similarities and differences between classical and quantum mechanical kinematics. Experimental studies of rotation in single, free superfluid $^4$He droplets have recently become feasible with the advent of x-ray free electron lasers (XFELs) and intense extreme ultraviolet (XUV) sources [7-10]. Ultrafast diffraction experiments have established the existence of oblate axisymmetric and triaxial prolate superfluid droplets with shapes that are also found in classical droplets [1-4]. However, previous experiments were not able to independently probe the angular momentum and angular velocity of rotating droplets, which is imperative for understanding their underlying laws of motion.

The shapes of classical droplets undergoing RBR follow a universal stability diagram in terms of reduced angular momentum $\Lambda_{RBR}$ and reduced angular velocity $\Omega_{RBR}$, as shown in Fig. 1 [1-4]. The reduced quantities are linked to the angular momentum $L_{RBR}$ and angular velocity $\omega_{RBR}$ according to

$$\Lambda_{RBR} = \frac{1}{\sqrt{8 \cdot \sigma \cdot \rho \cdot R^7}} L_{RBR} \qquad (1)$$

$$\Omega_{RBR} = \sqrt{\frac{\rho \cdot R^3}{8 \cdot \sigma}} \, \omega_{RBR} \, . \qquad (2)$$

Here, $\sigma$ is the surface tension of the liquid; $\rho$ its mass density; and $R$ the radius of a spherical droplet with the same volume as the deformed droplet. With increasing $\Lambda_{RBR}$, the droplets become more oblate. At $\Lambda_{RBR} > 1.2$, however, they assume prolate shapes such as pseudo-triaxial ellipsoids, capsules and dumbbells. It is unknown whether this diagram also applies to superfluid droplets. We address this fundamental question by simultaneously determining the angular momentum, $L$, the angular velocity, $\omega$, and the corresponding shapes of the rotating droplets via imaging both the shapes of $^4$He droplets and the vortex configurations they contain.

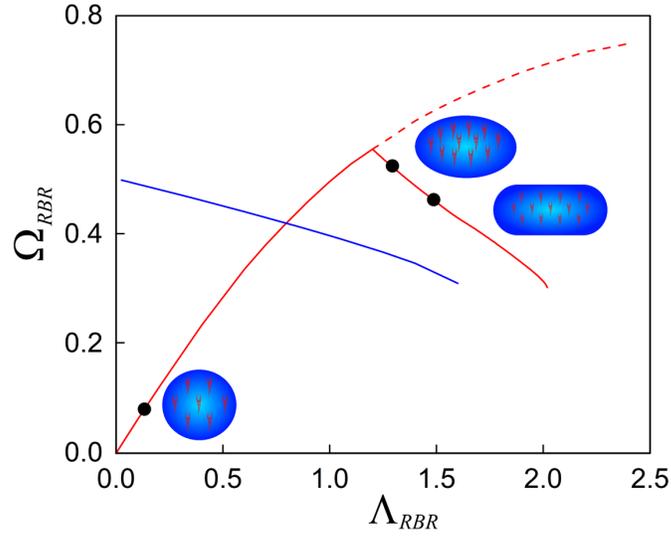

**Figure 1.** Stability diagram for classical droplets executing RBR. Red solid traces indicate stable shapes for specific combinations of reduced angular velocity $\Omega_{RBR}$ and reduced angular momentum $\Lambda_{RBR}$ [3,4]. The left branch corresponds to oblate axisymmetric shapes, the right branch to prolate shapes. The unstable portion of the branch is indicated by a dashed curve. The blue curve corresponds to droplets rotating solely through capillary wave motion [11]. Black circles mark the locations of classical droplets with the same aspect ratios as the superfluid droplets studied in this work.

Rotation of prolate superfluid droplets is interesting because it can be accomplished by capillary waves, which decay rapidly in classical droplets due to viscosity-induced energy dissipation but may, in principle, be sustained indefinitely in superfluids. Capillary waves in levitating helium droplets have been investigated experimentally and theoretically [12,13]. Travelling capillary waves give rise to a new branch in the stability diagram as shown by the blue line in Fig. 1 [11]. Prolate superfluid droplets likely contain both capillary waves and vortices, whose combined action determines the droplet shape. The arrangement of vortices in rotating, non-axisymmetric systems represents a general problem that has been discussed theoretically in connection with liquid helium [11,14] and dilute Bose-Einstein condensates (BECs) [15-18]. The observation, however, of vortices in rotating systems with large asymmetries remains challenging [19].

Here, we posit that in axisymmetric, oblate superfluid droplets, angular momentum originates solely from quantized vortices that arrange in a triangular lattice [7,20] similar to those previously observed in BECs [5,6,21]. In contrast, the angular momentum in prolate superfluid droplets has significant contributions from both vortices and capillary waves. We find that in capsule-shaped droplets, vortices form a triangular lattice, whereas they are arranged along elliptical contours in ellipsoidal droplets. The combined action of the vortex lattice and the capillary waves yields droplet shapes close to those in isochoric classical droplets rotating at the same angular velocity.

Experiments have been performed using the LAMP endstation at the Atomic, Molecular and Optical (AMO) beamline of the Linac Coherent Light Source (LCLS) XFEL [22,23]. Sub-micrometer sized helium droplets containing $N_{He} = 10^9$-$10^{11}$ atoms are produced via fragmentation of liquid $^4$He expanded into vacuum through a 5 μm nozzle at a temperature of 4.5 K and a backing pressure of 20 bar [7,24,25]. The droplets evaporatively cool in vacuum to T = 0.37 K [25,26], becoming superfluid at T ≈ 2.17 K. In order to visualize vortices, the droplets are doped with Xe atoms, which cluster along the vortex cores [7,27-30]. The XFEL is operated at 120 Hz with a photon energy of 1.5 keV (λ = 0.826 nm) and a pulse width of < 180 fs. The ultrashort duration and high intensity of the x-ray pulses enables the instantaneous capture of the structure of individual droplets and the vortex arrays they contain. Diffraction images are recorded on a pn-charge-coupled device (pn-CCD) detector that is centered along the XFEL beam axis. Since the diffraction images span only small scattering angles, they predominantly contain column density information of the scattering object in the direction perpendicular to the detector plane.

Diffraction patterns from doped droplets with slightly oblate axisymmetric, triaxial pseudo-ellipsoidal, and capsule shapes are presented in a logarithmic color scale in Fig. 2 A1, B1, and C1, respectively. Figure 2 A2-C2 shows the corresponding column densities of He (blue) and Xe (red/yellow) obtained from the diffraction patterns using the droplet coherent diffractive imaging (DCDI) algorithm [28,31], see SM for details [32]. The diffraction pattern in Fig. 2 A1 exhibits a circular ring structure close to the center and a speckled pattern in the outer region, which is due to scattering off the He droplet and embedded Xe clusters, respectively. From the inter-ring spacing at different azimuthal angles [7,27,31], the projection of the droplet contour onto the detector plane is found to be circular with a radius of 308±6 nm. As seen in Fig. 2 A2, the droplet contains 12 vortex cores, visualized by Xe filaments. The filaments are aligned along the droplet rotational axis and extend between opposite surface points. The filaments appear much shorter than the droplet diameter, indicating that the rotational axis is tilted with respect to the x-ray beam by ≈ 0.3 rad.

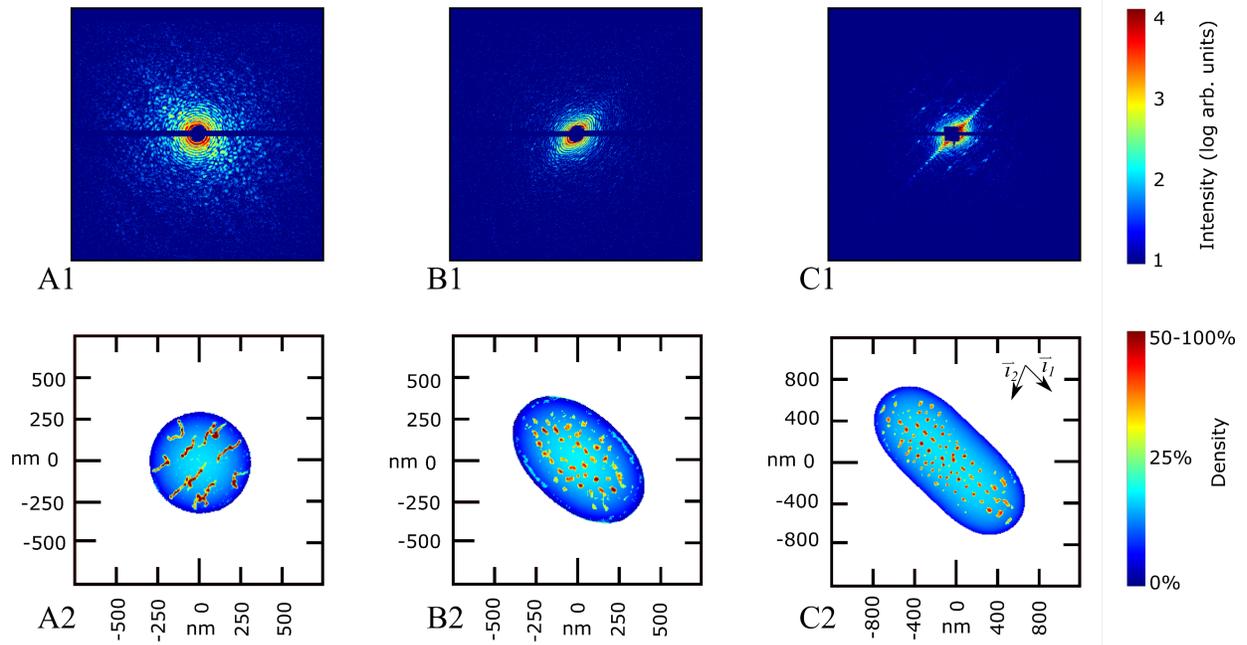

**Figure 2.** Diffraction patterns from Xe-doped droplets with various shapes: (A1) axisymmetric, nearly spherical, (B1) triaxial pseudo-ellipsoidal and (C1) capsule-shaped. The blank horizontal stripe in A1-C1 results from the gap between the upper and lower detector panels. Panels A2-C2 show column densities retrieved via the DCDI algorithm. The basis vectors of the vortex lattice in C2 are shown in the upper right corner of the panel. The vector lengths are exaggerated for better visibility.

Figures 2 B1 and C1 exhibit elongated diffraction contours, with Fig. 2 C1 also featuring a pronounced streak and a regular pattern of Bragg spots. Most noticeably, two parallel lines of four Bragg spots each are arranged symmetrically with respect to the streak. We assign the Bragg spots to x-ray scattering off a lattice of vortex-bound Xe clusters [7]. Figure 2 C2 shows that the droplet contains ~50 vortices arranged in four rows. The Xe clusters in Fig. 2 C2 appear as compact spots, although they are likely filaments as in Fig. 2 A2. Therefore, the filaments must be aligned

predominantly perpendicular to the detector plane, indicating a negligible tilt of the droplet's angular momentum with respect to the x-ray beam. In comparison, the droplet in Fig. 2 B2 has a pseudo-ellipsoidal shape and the pattern of vortices within resemble the droplet's outer contour. The Xe clusters show some elongation along the long axis, indicating a tilt of ≤ 0.15 rad around the droplet's short axis. Because the angular momentum of the droplets in panels B and C is aligned approximately along the x-ray beam, the droplet contours in panels B2 and C2 yield the actual values of the major and intermediate half axes, *a* and *b*, as well as the droplet aspect ratio: AR=*a*/*b*. The minor half-axis *c*, parallel to the rotational axis, cannot be directly determined.

Figures 2 A2, B2 and C2 showcase the evolution of vortex arrangements with changing droplet shapes. In the spheroidal droplet in Fig. 2 A2, the vortices are distributed across the entire volume and at regular distances from each other. In the prolate triaxial droplet in Fig. 2 B2, the vortices occupy only the central part of the droplet, leaving a ~100 nm wide boundary region devoid of vortices. In the capsule-shaped droplet in Fig. 2 C2, the vortices are organized in four rows along the droplet's long axis and the vortex-free boundary region is ~150 nm wide. The vortex lattice is reciprocal to that for the Bragg spots and is defined as

$$\vec{R} = m_1 r_1 \vec{\iota}_1 + m_2 r_2 \vec{\iota}_2, \qquad (3)$$

in which $m_1$ and $m_2$ are integers, $r_1$ = 111 nm, $r_2$ = 100 nm, and $\vec{\iota}_1$ and $\vec{\iota}_2$ are unit vectors with $\vec{\iota}_1$ perpendicular to the streak and $\vec{\iota}_2$ oriented at a 67° angle relative to $\vec{\iota}_1$ as illustrated in Fig. 2 C2. The vortex lattice is triangular, but not equilateral as in the idealized 2D case [5] and as previously observed in axisymmetric rotating He droplets [7] and BECs [5,6,21]. These observations suggest that the droplet shapes and vortex patterns within are closely inter-related.

We compare the angular momentum and angular velocity in superfluid droplets in Fig. 2 B2, C2 to those of their classical counterparts with the same half axes. From the measured AR in Fig. 2 B2 and C2, we first obtain the values of the reduced quantities $\Lambda_{RBR}$ and $\Omega_{RBR}$ [3,4,8] for the corresponding classical rotating droplets and then calculate $L_{RBR}$ and $\omega_{RBR}$ according to eqs. (1) and (2). The angular momentum due to vortices, $L_{VORT}$, is obtained from the number of vortices, $N_V$, and their location within the droplet [33] as detailed in the SM [32]. The angular velocity of the vortex array is obtained from the average vorticity as [34]

$$\omega_{VORT} = \frac{n_v \kappa}{2} \qquad (4)$$

where $n_v$ is the areal number density of vortices and $\kappa = h/m_4$ is the quantum of circulation, derived from Planck's constant $h$ and the mass of the $^4$He atom $m_4$.

The angular velocity of the superfluid droplet is given by $\omega_{SF} = \omega_{VORT}$ because, at equilibrium, the vortex array must be stationary with respect to the droplet contour. The angular momentum due to capillary waves, $L_{CAP}$, is obtained as the product of the effective irrotational moment of inertia of the droplets multiplied by $\omega_{SF}$ [32,35]. Contributions of Xe tracers to the angular momentum are neglected because the total mass of the embedded Xe atoms in Fig. 2 is approximately a factor of 100 smaller than that of the He droplet itself.

| Quantity/droplet | A2 | B2 | C2 |
|---|---|---|---|
| $L_{RBR}/(N_{He} \times \hbar)$ | 4.8 | 50 | 73 |
| $\omega_{RBR}$, rad/s | $2.0 \times 10^6$ | $1.24 \times 10^7$ | $4.9 \times 10^6$ |
| $\omega_{SF}$, rad/s | $2.0 \times 10^6$ | $9.7 \times 10^6$ | $4.9 \times 10^6$ |
| $L_{VORT}/(N_{He} \times \hbar)$ | 4.8 | 31 | 30 |
| $L_{CAP}/(N_{He} \times \hbar)$ | 0 | 18 | 48 |
| $L_{SF} = L_{VORT} + L_{CAP}$ | 4.8 | 49 | 78 |

Table 1. Kinematic parameters for the droplets in Figure 2.

Table 1 lists the kinematic parameters of the rotating superfluid droplets in Fig 2. It shows that the total angular momentum in superfluid prolate droplets, $L_{SF}$, (B2, C2) has contributions from both $L_{VORT}$ and $L_{CAP}$. The $L_{CAP}/L_{VORT}$ ratio decreases with decreasing AR, ultimately meeting the condition $L_{CAP} = 0$ in axisymmetric droplets (A2). The analysis also reveals that, within the accuracy of our estimates of $L$ and $\omega$, $L_{SF} \approx L_{RBR}$ and $\omega_{SF} \approx \omega_{RBR}$ for classical and vortex-hosting superfluid droplets of similar shape. More generally, the results in Table 1 suggest that rotation of large, vortex-hosting superfluid droplets can be described in terms of a reduced angular velocity and a reduced angular momentum in a very similar fashion as the rotational motion of their classical counterparts. Details of the analysis and a comprehensive list of geometric and kinematic parameters are presented in the SM [32], which also documents some differences between classical and superfluid shapes having the same AR.

To rationalize the similarities between the shapes of classical and superfluid droplets, we may recall that, for a given $L$, RBR has the smallest kinetic energy. A vortex array in the central part of a droplet creates a region of liquid that effectively moves as in RBR and minimizes the required capillary engagement. Essentially, the prolate droplet attempts to approach RBR with available means, i.e., by combining the effect of vortices and capillary waves. It appears that for each $L$, there is an optimal partitioning into $L_{VORT}$ and $L_{CAP}$ that minimizes the total energy of the rotating droplet. States with larger or smaller numbers of vortices are metastable and appear in the regions in between the stability curves for capillary wave motion and axisymmetric droplets in Fig. 1.

To ascertain how the shape of a superfluid droplet changes with the number of vortices at constant angular momentum, we have carried out density functional theory (DFT) calculations [11,36,37]. Due to the high computational cost, a full 3D calculation can only be performed for small droplets, which cannot support a large number of vortices [11]. Therefore, we have studied a free-standing, deformable $^4$He cylinder rotating around its symmetry axis. The calculations are performed in the co-rotating frame [37]. The total angular momentum is fixed at $L_{SF}/N_{He}=7.83\hbar$ .

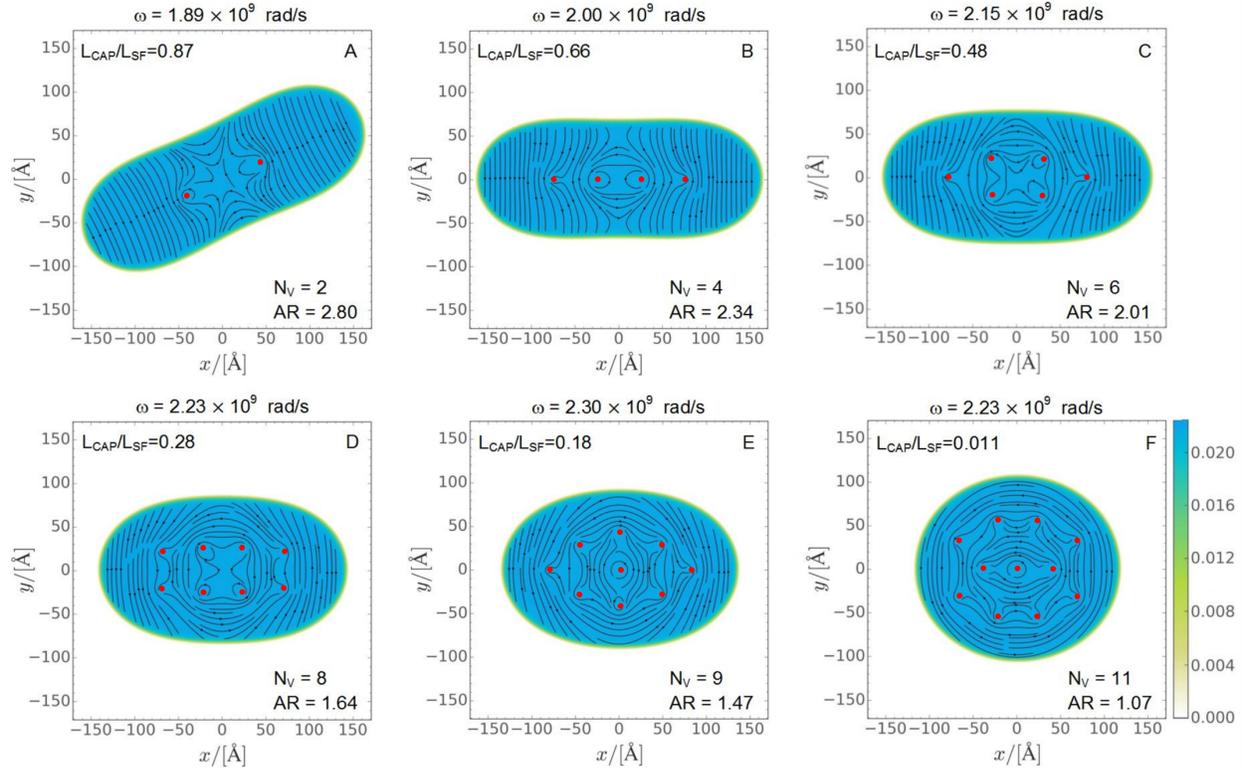

Figure 3. Calculated equilibrium density profiles of a deformable $^4$He cylinder rotating around its symmetry axis at fixed $L_{SF}/N_{He} = 7.83\hbar$ for different numbers of vortices. Streamlines are shown in black. The positions of the vortex cores are marked by red dots for better visualization. The color bar shows the density in units of Å$^{-3}$.

Figure 3A-F show the results of the calculations with different numbers of vortices, ranging from $N_V = 2$ to $N_V = 11$. The deformation of the cylinder increases for a decreasing number of vortices, as capillary waves contribute more angular momentum. We note that the energies of the configurations in Fig. 3 are rather similar; they differ less than 0.01 K per atom, with the $N_V = 6$ configuration (Fig. 3C) representing the global minimum. Each panel in Fig. 3 shows the computed $L_{CAP}/L_{SF}$ ratio and the cylinder's AR and $\omega$. In cylinders with relatively small AR = 1.1 and 1.5, the vortices are arranged into elliptical patterns, whereas they form elongated arrays for larger ARs, in agreement with the experimental observations in Fig. 2.

Each panel in Fig. 3 also shows the superfluid streamlines, illustrating the different irrotational velocity fields: the streamlines dominated by vortices wrap around the vortex cores, whereas those associated with capillary waves terminate at the surface. Figure 3 illustrates how the simultaneous effect of both vortices and capillary waves confers the appearance of a rotating rigid body to the motion of the superfluid.

Figure 3D shows an elongated droplet containing 8 vortices arranged in a pattern reminiscent of a distorted square lattice with an average nearest-neighbor distance of $d = 4.7$ nm and an areal vortex density of $1/d^2 = 4.5 \times 10^{16}$ m$^{-2}$. According to eq. (4), the angular velocity is $\omega = 2.26 \times 10^9$ rad/s in excellent agreement with the DFT result of $\omega = 2.23 \times 10^9$ rad/s. Thus, the results of the DFT calculations lend further support for using eq. (4) to determine ω of rotating He droplets from the finite vortex arrays they contain.

This combined experimental and theoretical study provides a direct comparison between the shapes, angular velocities, and angular momenta of rotating classical and superfluid droplets. The results show that the equilibrium figures of superfluid droplets hosting vortices may be described in a similar fashion as those of classical, viscous droplets by a series of oblate and prolate shapes that evolve along curves of stability as a function of reduced angular momentum Λ and reduced angular velocity Ω. In axisymmetric oblate superfluid droplets, all angular momentum is stored in triangular lattices of quantum vortices that extend throughout the entire droplet volume. However, in prolate droplets, capillary waves contribute an increasing and, in some cases, dominant amount of angular momentum. It appears that prolate superfluid droplets approach the classical equilibrium figures through a combination of vortices and capillary motion, whereby the vortex arrangements are restricted to the central region of the droplet.

Future experiments at high-repetition rate ultrafast x-ray light sources such as the European XFEL and LCLS-II will explore a larger phase space in terms of Λ, Ω, and R. More advanced theoretical work is required to study angular velocities and angular momenta with higher accuracy and to derive a more detailed stability diagram of superfluid droplets from the measurements. Furthermore, probing smaller droplet sizes will reveal possible finite size effects.

**Acknowledgements**

This work was supported by the NSF Grant No DMR-1701077 (A.F.V.), by the U.S. Department of Energy, Office of Basic Energy Sciences, (DOE, OBES) Chemical Sciences, Geosciences and Biosciences Division, through Contract No. DE-AC02-05CH11231 (C.Ba., C.A.S., J.M., B.W.T., O.G.) as well as DE-AC02-06CH11357 (C.Bo.), DE-AC02-76SF00515 (C.Bo.), Université Fédérale Toulouse-Midi-Pyrénées throughout the "Chaires d'Attractivité 2014" Programme IMDYNHE (M.B.), and FIS2017-87801-P (AEI/FEDER, UE) (M.B., M.P.). Portions of this research were carried out at the Linac Coherent Light Source, a national user facility operated by Stanford University on behalf of the U.S. DOE, OBES under beam-time grant LP05: Superfluids far from equilibrium. We are grateful to John Bozek, Justin Kwok, Curtis Jones, Ken Ferguson, Sebastian Schorb, and Martin Seifrid for providing assistance during some experiments described in this paper.

# References

[1]  S. Chandrasekhar, *The Stability of a Rotating Liquid Drop*. Proc. Roy. Soc. London A **186**, 1 (1965).
[2]  S. Cohen, F. Plasil, and W. J. Swiatecki, *Equilibrium Configurations of Rotating Charged or Gravitating Liquid Masses with Surface-Tension .2.* Annals of Physics **82**, 557 (1974).
[3]  R. A. Brown and L. E. Scriven, *The Shape and Stability of Rotating Liquid-Drops*. Proc. R. Soc. London, Ser. A **371**, 331 (1980).
[4]  S. L. Butler, M. R. Stauffer, G. Sinha, A. Lilly, and R. J. Spiteri, *The Shape Distribution of Splash-Form Tektites Predicted by Numerical Simulations of Rotating Fluid Drops*. J. Fluid Mech. **667**, 358 (2011).
[5]  L. Pitaevskii and S. Stringari, *Bose-Einstein Condensation and Superfluidity*. Bose-Einstein Condensation and Superfluidity **164**, Oxford University Press (2016).
[6]  A. L. Fetter, *Rotating Trapped Bose-Einstein Condensates*. Rev. Mod. Phys. **81**, 647 (2009).
[7]  L. F. Gomez *et al.*, *Shapes and Vorticities of Superfluid Helium Nanodroplets*. Science **345**, 906 (2014).
[8]  C. Bernando *et al.*, *Shapes of Rotating Superfluid Helium Nanodroplets*. Phys. Rev. B **95**, 064510 (2017).
[9]  D. Rupp *et al.*, *Coherent Diffractive Imaging of Single Helium Nanodroplets with a High Harmonic Generation Source*. Nature Comm. **8**, 493 (2017).
[10]  B. Langbehn *et al.*, *Three-Dimensional Shapes of Spinning Helium Nanodroplets*. Phys. Rev. Lett. **121**, 255301 (2018).
[11]  F. Ancilotto, M. Barranco, and M. Pi, *Spinning Superfluid He-4 Nanodroplets*. Phys. Rev. B **97**, 184515 (2018).
[12]  D. L. Whitaker, M. A. Weilert, C. L. Vicente, H. J. Maris, and G. M. Seidel, *Oscillations of Charged Helium II Drops*. J. Low Temp. Phys. **110**, 173 (1998).
[13]  L. Childress, M. P. Schmidt, A. D. Kashkanova, C. D. Brown, G. I. Harris, A. Aiello, F. Marquardt, and J. G. E. Harris, *Cavity Optomechanics in a Levitated Helium Drop*. Phys. Rev. A **96**, 063842 (2017).
[14]  A. L. Fetter, *Vortex Nucleation in Deformed Rotating Cylinders*. J. Low Temp. Phys. **16**, 533 (1974).
[15]  M. O. Oktel, *Vortex Lattice of a Bose-Einstein Condensate in a Rotating Anisotropic Trap*. Phys. Rev. A **69**, 023618 (2004).
[16]  S. Sinha and G. V. Shlyapnikov, *Two-Dimensional Bose-Einstein Condensate Under Extreme Rotation*. Phys. Rev. Lett. **94**, 150401 (2005).
[17]  P. Sanchez-Lotero and J. J. Palacios, *Vortices in a Rotating Bose-Einstein Condensate Under Extreme Elongation*. Phys. Rev. A **72**, 043613 (2005).
[18]  N. Lo Gullo, T. Busch, and M. Paternostro, *Structural Change of Vortex Patterns in Anisotropic Bose-Einstein Condensates*. Phys. Rev. A **83**, 063632 (2011).
[19]  K. Deconde and R. E. Packard, *Measurement of Equilibrium Critical Velocities for Vortex Formation in Superfluid-Helium*. Phys. Rev. Lett. **35**, 732 (1975).

[20]	F. Ancilotto, M. Pi, and M. Barranco, *Vortex Arrays in Nanoscopic Superfluid Helium Droplets*. Phys. Rev. B **91**, 100503(R) (2015).
[21]	J. R. Abo-Shaeer, C. Raman, J. M. Vogels, and W. Ketterle, *Observation of Vortex Lattices in Bose-Einstein Condensates*. Science **292**, 476 (2001).
[22]	L. Struder *et al.*, *Large-format, High-speed, X-ray pnCCDs Combined with Electron and Ion Imaging Spectrometers in a Multipurpose Chamber for Experiments at 4th Generation Light Sources*. Nucl. Instrum. Methods Phys. Res. A **614**, 483 (2010).
[23]	K. R. Ferguson *et al.*, *The Atomic, Molecular and Optical Science instrument at the Linac Coherent Light Source*. J Synchrotron Radiat **22**, 492 (2015).
[24]	L. F. Gomez, E. Loginov, R. Sliter, and A. F. Vilesov, *Sizes of Large He Droplets*. J. Chem. Phys. **135**, 154201 (2011).
[25]	R. M. P. Tanyag, C. F. Jones, C. Bernando, D. Verma, S. M. O. O'Connell, and A. F. Vilesov, in *Cold Chemistry: Molecular Scattering and Reactivity Near Absolute Zero*, edited by A. Osterwalder, and O. Dulieu (Royal Society of Chemistry, Cambridge, 2018), p. 389.
[26]	M. Hartmann, R. E. Miller, J. P. Toennies, and A. Vilesov, *Rotationally Resolved Spectroscopy of SF6 in Liquid-Helium Clusters - a Molecular Probe of Cluster Temperature*. Phys. Rev. Lett. **75**, 1566 (1995).
[27]	C. F. Jones *et al.*, *Coupled Motion of Xe Clusters and Quantum Vortices in He Nanodroplets*. Phys. Rev. B **93**, 180510 (2016).
[28]	O. Gessner and A. F. Vilesov, *Imaging Quantum Vortices in Superfluid Helium Droplets*. Annu. Rev. Phys. Chem. **70**, 173 (2019).
[29]	F. Dalfovo, R. Mayol, M. Pi, and M. Barranco, *Pinning of Quantized Vortices in Helium Drops by Dopant Atoms and Molecules*. Phys. Rev. Lett. **85**, 1028 (2000).
[30]	G. P. Bewley, D. P. Lathrop, and K. R. Sreenivasan, *Superfluid Helium - Visualization of Quantized Vortices*. Nature **441**, 588 (2006).
[31]	R. M. P. Tanyag *et al.*, *Communication: X-ray Coherent Diffractive Imaging by Immersion in Nanodroplets*. Struct. Dyn. **2**, 051102 (2015).
[32]	See Supplemental Materials at //link// for additional information on the determination of the droplet parameters, some deviations between the classical and superfluid droplet shapes and details of DFT calculations, which includes references [7,8,11,27,31,33-37].
[33]	S. T. Nam, G. H. Bauer, and R. J. Donnelly, *Vortex Patterns in a Freely Rotating Superfluid*. J. Korean Phys. Soc. **29**, 755 (1996).
[34]	R. P. Feynman, *Chapter II Application of Quantum Mechanics to Liquid Helium*, C. J. Gorter Ed., North-Holland Publishing Company: Amsterdam, 1955: Vol. 2, p. 17.
[35]	F. Coppens, F. Ancilotto, M. Barranco, N. Halberstadt, and M. Pi, *Capture of Xe and Ar Atoms by Quantized Vortices in He-4 Nanodroplets*. Phys. Chem. Chem. Phys **19**, 24805 (2017).
[36]	F. Ancilotto, M. Barranco, F. Coppens, J. Eloranta, N. Halberstadt, A. Hernando, D. Mateo, and M. Pi, *Density Functional Theory of Doped Superfluid Liquid Helium and Nanodroplets*. Int. Rev. Phys. Chem. **36**, 174512 (2017).
[37]	F. Ancilotto, M. Pi, and M. Barranco, *Vortex Arrays in a Rotating Superfluid He-4 Nanocylinder*. Phys. Rev. B **90**, 174512 (2014).

# Angular momentum in rotating superfluid droplets

# Supplemental Materials

Sean M. O'Connell[1], Rico Mayro P. Tanyag[1,2], Deepak Verma[1], Charles Bernando[3,4], Weiwu Pang[5], Camila Bacellar[6,7], Catherine A. Saladrigas[6,7], Johannes Mahl[6,8], Benjamin W. Toulson[6], Yoshiaki Kumagai[9], Peter Walter[10], Francesco Ancilotto[11,12,a)], Manuel Barranco[13,14,15], Marti Pi[13,14], Christoph Bostedt[9,16,17,18,19], Oliver Gessner[6,a)] and Andrey F. Vilesov[1,3,a)]

[1] Department of Chemistry, University of Southern California, Los Angeles, California 90089, USA
[2] Technische Universät Berlin, Institut für Optik und Atomare Physik, 10623 Berlin, Germany
[3] Department of Physics and Astronomy, University of Southern California, Los Angeles, California 90089, USA
[4] OVO (PT. Visionet Internasional), Lippo Kuningan 20th floor, Jl. HR Rasuna Said Kav. B-12, Setiabudi, Jakarta, 12940, Indonesia
[5] Department of Computer Science, University of Southern California, Los Angeles, California 90089, USA
[6] Chemical Sciences Division, Lawrence Berkeley National Laboratory, Berkeley, California 94720, USA
[7] Department of Chemistry, University of California Berkeley, Berkeley, California 94720, USA
[8] Department of Physics, University of Hamburg, 22761 Hamburg, Germany
[9] Argonne National Laboratory, 9700 South Cass Avenue B109, Lemont, Illinois 60439, USA
[10] Linac Coherent Light Source, SLAC National Accelerator Laboratory, Menlo Park, California, 94025, USA
[11] Dipartimento di Fisica e Astronomia and CNISM, Università di Padova, 35122 Padova, Italy
[12] CNR-IOM Democritos, 34136 Trieste, Italy
[13] Departament FQA, Universitat de Barcelona, Facultat de Física, 08028 Barcelona, Spain
[14] Institute of Nanoscience and Nanotechnology (IN2UB), Universitat de Barcelona, Barcelona, Spain
[15] Laboratoire des Collisions, Agrégats et Réactivité, IRSAMC, Université Toulouse 3, F-31062 Toulouse, France
[16] Department of Physics and Astronomy, Northwestern University, 2145 Sheridan Road, Evanston, Illinois 60208, USA
[17] Paul Scherrer Institut, 5232 Villigen – PSI, Switzerland
[18] LUXS Laboratory for Ultrafast X-ray Sciences, Institute of Chemical Sciences and Engineering, Lausanne, Switzerland
[19] École Polytechnique Fédérale de Lausanne (EPFL), CH-1015, Lausanne, Switzerland

a)Authors to whom correspondence should be addressed. Electronic addresses: francesco.ancilotto@pd.infn.it; ogessner@lbl.gov; and vilesov@usc.edu.

1. **Droplet Coherent Diffractive Imaging and Density Reconstructions**

Previously, it was shown that densities of clusters encapsulated in helium droplets can be reconstructed from the x-ray diffraction patterns using the so-called droplet coherent diffractive imaging algorithm (DCDI) [1]. The algorithm exploits the fact that the scattering amplitude from a doped He droplet can be calculated as the sum of the known amplitude from the He droplet itself and the unknown amplitude from the dopants. Since the experiments were performed at small scattering angles [1-4], the diffraction amplitude can be well approximated by the two-dimensional Fourier Transform of the column density of the scattering object in the direction perpendicular to the detector plane. Accordingly, the DCDI algorithm requires the column density of the droplet as an input. For a spheroidal droplet, the column density can be calculated analytically from the half axes of the droplet's projection on the detector plane. The half axes are obtained from the elliptical droplet diffraction contours as described previously [1,2,4]. The density is then normalized to fit the intensity of the diffraction signal from the droplet [1]. However, the shape of prolate droplets cannot be obtained from the half axes because their shapes are not given by any analytic expression.

Therefore, we apply a numerical determination of the column densities for prolate droplets. First, the contour of the droplet's projection onto the detector plane is determined. It is obtained via 2D inverse Fourier Transform (IFT) [4]. In Ref. [4], before applying IFT, the phase was assigned to each of the diffraction contours from the droplet, which is a tedious procedure. Therefore, in this work we employ the IFT of the square root of the diffraction intensity itself, which yields similar results in a less cumbersome manner. Because the input is positive throughout, it can be seen as a smooth positive offset function decreasing towards larger scattering angles with an oscillating component. It is easy to show that the period of oscillation is a factor of two smaller than that in the diffraction amplitude. Therefore, the obtained droplet contour is scaled down by a factor of two. Figure S1 shows the result of the application of the IFT to the square root of the diffraction intensity shown in Fig. 2 C1 of the main text.

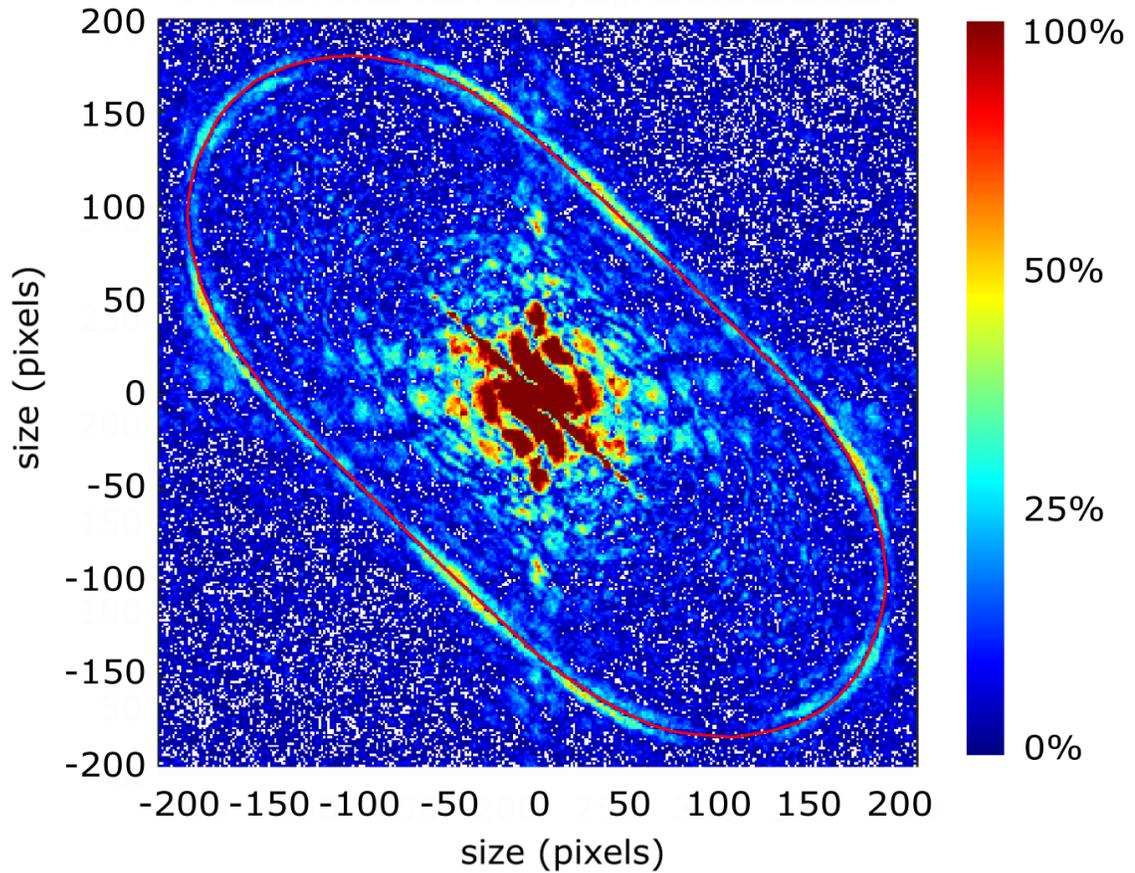

**Figure S1**: Droplet shape reconstruction from inverse Fourier Transform of the square root of the diffraction image in Figure 2 C1 in the main text. The result of the IFT is shown in a linear color scale. There are two contours of similar intensity with a superimposed red line in between, which marks the droplet contour. The signals inside the droplet are caused by a positive offset, Bragg peaks, and other features of the original diffraction as discussed in the text.

Figure S1 shows the contour of the droplet in light blue and yellow colors, the long axis of which is aligned diagonally. Intensities inside of the contour are due to the IFT of the offset function, Bragg spots and other low frequency features in the diffraction. Similar to previous work [4], the procedure yields multiple contours because some low-frequency signals from the droplet contour are not recorded due to the central hole in the detector. The true droplet contour is expected to coincide with the center of gravity of the modulus of these contours. Figure S1 shows a set of two contours of comparable amplitude. Therefore, the droplet's contour lies in between, as indicated by the red line. Once the droplet contour is obtained, the density of the droplet is filled in axisymmetrically with respect to the long axis of the contour. This density is an approximation, because the droplet *b*-axis perpendicular to the rotation axis is slightly larger than the *c*-axis parallel to the rotation axis as noted in the main text [4]. Therefore, the actual density distribution of the prolate droplet is not strictly axisymmetric with respect to the long axis.

## 2. Geometric and Kinematic Parameters of Reconstructed Droplets

Table S1. Geometric and kinematic parameters for the droplets in Figure 2 A2-C2 in main text

| Quantity/droplet | A2 | B2 | C2 |
| --- | --- | --- | --- |
| *a*, nm | 308 | 466 | 883 |
| *b*, nm | 308 | 290 | 389 |
| AR | 1.00 (exp)<br>1.007 (from $\Omega_{RBR}$) | 1.61 | 2.27 |
| $N_{He}$ | $2.7\times10^9$ | $3.1\times10^9$ | $1.45\times10^{10}$ |
| $\Lambda_{RBR}$ | 0.13 | 1.31 | 1.5 |
| $L_{RBR}/(N_{He}\times\hbar)$ | 4.8 | 50 | 73 |
| $\Omega_{RBR}$ | 0.08 | 0.52 | 0.46 |
| $\omega_{RBR}$, rad/s | $2.0\times10^6$ | $1.24\times10^7$ | $4.9\times10^6$ |
| $N_V$ | 12 | 50 | 50 |
| $S_V$, nm | 170 | 77 | 109 |
| $n_V$, m$^{-2}$ | $4.0\times10^{13}$ | $2.0\times10^{14}$ | $9.8\times10^{13}$ |
| $\omega_{SF}=\omega_{VORT}$, rad/s | $2.0\times10^6$ | $9.7\times10^6$ | $4.9\times10^6$ |
| $L_{VORT}/(N_{He}\times\hbar)$ | 4.8 ($R_B = R$) | 31 ($R_B = .75 R$) | 30 ($R_B = .78 R$) |
| $L_{CAP}/(N_{He}\times\hbar)$ | 0 | 18 | 48 |
| $L_{SF}=L_{CAP}+L_{VORT}/(N_{He}\times\hbar)$ | 4.8 | 49 | 78 |

Table S1 shows the geometric and kinematic parameters of for the droplets in Fig. 2 A2-C2 of the main text. Shown in the first three rows are the long and short half-axis of the droplet and the aspect ratios, *a*, *b* and AR=*a*/*b*, respectively, which were obtained from the diffraction contour patterns [1,2,4]. The number of atoms in the droplet, $N_{He}$, is determined from the volume of a classical droplet having same *a* and *b* and the number density of liquid helium ($n_{He} = 2.18\times10^{28}$ atoms· m$^{-3}$), see Fig. 8 in Bernando *et al.* [4]. The values of reduced angular momentum, $\Lambda_{RBR}$, and reduced angular velocity, $\Omega_{RBR}$, are derived from AR, assuming the shapes of rigid body rotation (RBR), see Fig. 8 in Bernando *et al.*[4] The angular momentum, $L_{RBR}$, and angular velocity, $\omega_{RBR}$, are deduced using equations 1 and 2 of the main text, respectively, which are identical to eqs. (S1,S2).

$$\Lambda_{RBR} = \frac{1}{\sqrt{8 \cdot \sigma \cdot \rho \cdot R^7}} L_{RBR} \qquad (S1)$$

$$\Omega_{RBR} = \sqrt{\frac{\rho \cdot R^3}{8 \cdot \sigma}} \omega_{RBR} \qquad (S2)$$

$L_{RBR}$ is expressed in units of $\hbar$ per He atom, and $\omega_{RBR}$ is expressed in radians per second. For the droplet in A2, $\Lambda_{RBR}$ and $\Omega_{RBR}$ cannot be obtained in the same way, because its AR is unity within the experimental accuracy. Therefore, in this case, the results are based on $\omega_{SF}$, which is set to $\omega_{RBR}$ from which $\Lambda_{RBR}$ and $\Omega_{RBR}$ are deduced using eqs. S1 and S2. The angular velocity of the lattice is given as $\omega_{SF} = \frac{n_v \kappa}{2}$ [5], in which $n_v$ is the vortex density described below, and $\kappa = \frac{h}{m_4}$ is the quantum of circulation. Here $h$ is Planck's constant and $m_4$ is the mass of the $^4$He atom. For A2, it is assumed that vortices are uniformly distributed throughout the spherical droplet, thus, $n_v = \frac{N_v}{\pi R^2}$. For droplets B2 and C2, a triangular lattice of vortices is assumed, so $n_v = \frac{2}{\sqrt{3}} S_v^{-2}$, where $S_V$ is the average nearest neighbor distance between vortices. For A2, $n_V$ is used to calculate the $S_V$, assuming a uniform distribution of vortex cores inside the spherical droplet. For B2, $S_V$ was estimated from the spacing between vortices along the left part of the vortex lattice. In C2, $S_V$ is derived from the vortex positions in a lattice as described by eq. (3) of the main text. In droplets A2 and C2, in Fig. 2 of the main text, the number of vortices $N_V$ is found by direct count, and in B2, $N_V$ is approximated as the product of the $n_V$ and the area filled with vortices, which was estimated to be an ellipse with half axes of 360 nm and 218 nm.

The expression for angular momentum of $N_V$ linear vortices in spherical droplets was described by Nam *et al.*[6] Here, we use their expression:

$$L_{SF} = \frac{2\rho\kappa}{3} \sum_{i=1}^{N_v} (R - r_i)^{\frac{3}{2}}, \qquad (S3)$$

in which $R$ is the radius of the droplet, $r_i$ is the displacement of the *i*-vortex from the center of the droplet, $\rho$ is the mass density of superfluid helium, and $\kappa$ is the quantum of circulation. For this analysis, it is assumed that vortices with constant areal density fill a circular region with radius $R_B$ around the droplet center so that there is a boundary region of the droplet devoid of vortices between $R_B$ and $R$. In the limit of large numbers of vortices, the summation in eq. (S3) can be replaced by integration. Integration yields the angular momentum due to vortices in units of $\hbar$ per He atom as:

$$\frac{L_{SF}}{N_{He} \cdot \hbar} = \frac{2N_V}{5} \frac{R^5 - (R^2 - R_B^2)^{\frac{5}{2}}}{R^3 \cdot R_B^2} \qquad (S4)$$

Here, we use eq. (S4) for a sphere to approximate non-spherical solutions for droplets B2 and C2. The value of $R_B$ is estimated by measuring the area filled with vortices, $A_V$, dividing this by the area of the droplet cross section, $A_D$, and taking the square root of the result as: $R_B = R \cdot \sqrt{\frac{A_V}{A_D}}$. The angular momentum of capillary waves in a triaxial ellipsoid rotating around the z-axis is given as [7]:

$$\frac{L_{CAP}}{N_{He} \cdot \hbar} = \omega_{SF} m_4 \left( \frac{(\langle x^2 \rangle - \langle y^2 \rangle)^2}{\langle x^2 \rangle + \langle y^2 \rangle} \right). \qquad (S5)$$

The estimated values of $L_{SF}$ contain some error due to uncertainties in determining $L_{VORT}$ and $L_{CAP}$. One is associated with the use of eq. (S4) which is strictly only valid for spherical droplets. No corresponding analytic expression is available for droplets of general shape. In addition, $L_{CAP}$ is calculated using an irrotational moment of inertia for a liquid constrained to an ellipsoid [7], whereas the actual droplet shape is not strictly ellipsoidal. Furthermore, the values for $\omega_{SF}$ are obtained from Feynman's equation, $\omega_{SF} = \frac{n_v \kappa}{2}$, which is valid for large 2D arrays, whereas this work considers finite 3D vortex arrays within droplets. We note, however, that the perfect match of $\omega_{SF}$ and $\omega_{RBR}$ in Fig. 2 C2, where $n_V$ can be determined with high accuracy, as well as the results of the DFT calculations described in the main text, indicate that the application of Feynman's equation is likely accurate.

### 3. Differences between the shapes of classical and superfluid prolate droplets

Upon detailed inspection, the results in Fig. 2 C of the main text indicate some deviation between the calculated RBR and measured superfluid shapes. At $\Lambda_{RBR} = 1.5$ as in Fig. 2 C2, the classical shape has a weak depression along the *b* - axis, which gives rise to additional nodal structures in the diffraction signal that appear as multiple, radially emanating streaks (see SM of Bernando et al. [4]). The diffraction pattern in Fig.2 C1 does not exhibit any such nodes, which may indicate that the presence of the vortex array and capillary waves leads to a stabilization of the straight outline of the droplet. Moreover, the diffraction maxima along the streak are not oriented perpendicular to the streak axis, as is the case for RBR shapes (see SM from Bernando et al. [4]). Instead, the pattern is tilted counterclockwise. This tilt likely indicates that the two sides of the droplet are not exactly parallel but include a small angle on the order of a few degrees. Thus, the droplet in Fig 2 C2 is likely non-centrosymmetric, as is the case in classical droplets, which may be another indication of the interaction of the vortex lattice with the droplet shape. This wedging effect cannot be seen in the droplet contour in Fig. 2 C2 obtained from the Fourier Transform of the diffraction intensity, [4] which exclusively leads to axisymmetric shapes.

## 4. Density Functional calculations for rotating, deformable $^4$He cylinders

We consider a self-bound superfluid $^4$He cylinder rotating around its symmetry z-axis with a constant angular velocity $\omega$. The details of the calculations are described in Ref. [8], where vortices in axisymmetric rotating cylinders were considered. Here, we have extended that work to select configurations with the vortex-hosting cylinder allowed to deform into shapes with non-circular cross sections. We have assumed in our calculation a uniform density along the z-direction, which implies that the resulting vortices always remain linear. A complex order parameter, $\Psi$, is found to represent the superfluid helium state at zero temperature [8,9] whose square modulus is the atomic density. The DFT equation is formulated in a rotating frame-of-reference with constant angular velocity ω (co-rotating frame), which follows from the variational minimization of the energy, and is solved looking for stationary solutions. Vortex structures and positions are optimized during the functional minimization, after having been phase-imprinted, as described in Refs. [8,9]. Simultaneously, we allow for deviations from the circular cross section ("prolate" rotating cylinders) using the same method as previously applied to prolate spinning droplets [10], i.e. solve the variational equation by imposing a given value for the angular momentum per atom $L_z$ and iteratively find the associated value of the rotational frequency $\omega$. Classically, such fixed $L_z$ calculations correspond to torque-free drops with an initially prescribed rotation ("isolated drops") and result in stable prolate configurations like the ones shown in Fig. 3 in the main text.

**References:**

[1] R. M. P. Tanyag et al., *Communication: X-ray Coherent Diffractive Imaging by Immersion in Nanodroplets*. Struct. Dyn. **2**, 051102 (2015).
[2] L. F. Gomez et al., *Shapes and Vorticities of Superfluid Helium Nanodroplets*. Science **345**, 906 (2014).
[3] C. F. Jones et al., *Coupled Motion of Xe Clusters and Quantum Vortices in He Nanodroplets*. Phys. Rev. B **93**, 180510 (2016).
[4] C. Bernando et al., *Shapes of Rotating Superfluid Helium Nanodroplets*. Phys. Rev. B **95**, 064510 (2017).
[5] R. P. Feynman, *Chapter II Application of Quantum Mechanics to Liquid Helium*, C. J. Gorter Ed., North-Holland Publishing Company: Amsterdam, 1955: Vol. 2, p. 17.
[6] S. T. Nam, G. H. Bauer, and R. J. Donnelly, *Vortex Patterns in a Freely Rotating Superfluid*. J. Korean Phys. Soc. **29**, 755 (1996).
[7] F. Coppens, F. Ancilotto, M. Barranco, N. Halberstadt, and M. Pi, *Capture of Xe and Ar Atoms by Quantized Vortices in He-4 Nanodroplets*. Phys. Chem. Chem. Phys **19**, 24805 (2017).
[8] F. Ancilotto, M. Pi, and M. Barranco, *Vortex Arrays in a Rotating Superfluid He-4 Nanocylinder*. Phys. Rev. B **90**, 174512 (2014).
[9] F. Ancilotto, M. Barranco, F. Coppens, J. Eloranta, N. Halberstadt, A. Hernando, D. Mateo, and M. Pi, *Density Functional Theory of Doped Superfluid Liquid Helium and Nanodroplets*. Int. Rev. Phys. Chem. **36**, 174512 (2017).
[10] F. Ancilotto, M. Barranco, and M. Pi, *Spinning Superfluid He-4 Nanodroplets*. Phys. Rev. B **97**, 184515 (2018).